Controlling electron projectile coherence effects using twisted electrons

A. L. Harris*

Physics Department, Illinois State University, Normal, IL, USA 61790

**Abstract**

In traditional scattering theory, the incident projectile is assumed to have an infinite coherence length. However, over the last decade, experimental and theoretical studies of collisions using heavy ion projectiles have shown that this assumption is not always valid. This has led to a growing number of studies that specifically examined the effects of the projectile's coherence length on collision cross sections. These studies have used heavy ion projectiles because they offer a straight-forward method to control the projectile's coherence length through its momentum, and using these techniques, it has been demonstrated that the projectile's coherence length alters the cross sections. In contrast, it is widely presumed that the coherence length of an electron projectile is always sufficiently large that any effects on the cross sections can be safely neglected. We show that, contrary to this prevailing opinion, coherence effects are observable for electron projectiles and they can be controlled. We calculate triple differential cross sections (TDCSs) for ionization of $H_2^+$ using twisted electron projectiles in the form of Laguerre-Gauss and Bessel electrons. Effects of the projectile's coherence length are observed through the presence or absence of two-slit interference features in the TDCSs. When the electron projectile's coherence length is large, ionization occurs from either nuclear center of the molecule, and two-slit interference features are visible in the TDCSs. In contrast, when the projectile's coherence length is small, ionization occurs from only one nuclear center and the TDCSs resemble those for ionization of atomic hydrogen. We demonstrate that the intrinsic parameters of the vortex projectiles, such as beam waist and opening angle, can be used to control the coherence length of electron projectiles.

**I. Introduction**

Since the early days of quantum mechanics, charged particle collisions have been a valuable tool to probe the electronic structure of atoms and molecules and elucidate few-body Coulomb interactions. The back-and-forth comparison between experimental measurements and theoretical models has led to sufficiently good agreement that our understanding of many charged particle collisions with simple targets was considered complete [1–6]. However, about a decade ago, measurements of ionization of helium by heavy ion projectiles exposed unexpected discrepancies between established theories and state-of-the-art measurements [7]. In the years since the experiments of [7], a variety of explanations for these discrepancies have been proposed, including, for example, experimental resolution [8], inaccurate theoretical treatment of the projectile-target interaction [9–12], and screening of the target nucleus by inactive electrons [13]. The current leading explanation in the literature is that the projectile's transverse coherence length, a feature typically assumed to be infinite in most quantum mechanical scattering theories, is in fact finite and must be considered when comparing theoretical results with experimental data [14–23]. Several works have demonstrated control of projectile coherence length [14–20,23–31] through experimental design. In particular, projectiles with small coherence length, and particularly those whose coherence length is similar in size or smaller than the target width, can lead to significant alterations of the collision cross sections [15,17]. These effects went unnoticed for many decades because agreement between experiment and theory was generally quite good (e.g. [32,33]) for total and single differential cross sections at small perturbation parameters (ratio of projectile charge to speed) [7,8,34–37]. However, as the availability of fully differential cross section measurements expanded, it opened the door to more rigorous test of theory [7,32,34,38–42], revealing gaps in our understanding and insights into the possible role of projectile coherence length.

*alharri@ilstu.edu

To date, most studies of projectile coherence length have used heavy ion projectiles due to their small deBroglie wavelength, and thus small transverse coherence length. In these cases, the transverse coherence length was directly manipulated by changing the projectile's momentum through either the energy or ion type (i.e. mass) [14–17,24], allowing for the effects of the coherence length on collision cross sections to be observed. For electron projectiles, controlling the coherence length through the momentum is more difficult because of their small mass. Even at large velocities, the electron's wavelength remains large, which leads to a coherence length that is typically greater than the target width, preventing coherence effects from appearing in the cross sections.

Here, we demonstrate a means to study and control the transverse coherence length of electron projectiles on a scale where its effects are clearly observable. We calculate triple differential cross sections (TDCSs) for ionization of $H_2^+$ using electron vortex projectiles (also called twisted electrons). We show that the intrinsic parameters of the vortex projectiles, such as beam waist and opening angle, can be used to control the coherence length and localization of the impinging projectile. The projectile is determined to be coherent or incoherent relative to the size of the target molecule by the presence or absence of interference features in the TDCSs.

In the $H_2^+$ molecule, the target electron density is localized around two nuclear centers. When the incident projectile wave packet is wide enough to perturb the target wave function at both centers (i.e. coherent), the ionization amplitudes from the two centers can interference. This yields an interference pattern in the TDCS in analogy with the Young's double slit interference experiment using light [43–47]. Thus, the TDCSs are expected to show signs of interference effects when the projectile's transverse coherence length is equal to, or larger than, the internuclear separation.

If the incident projectile is incoherent, its transverse coherence length is small, and it is only able to perturb the wave packet localized around one of the nuclear centers. In this case, the TDCS will resemble that of ionization from a single hydrogen atom and no interference features will be observed. Therefore, the shape of the TDCSs is expected to reveal the coherent nature of the incident projectile wave packet relative to the size of the target molecule.

To date, most work with electron vortex projectiles in atomic and molecular collisions has focused on atomic targets [22,48–60] using Bessel and Airy electron beams and is exclusively theoretical. These studies were motivated by the experimental demonstration of these types of electron beams [61–64] and they have predicted that the use of electron vortex projectiles will significantly alter the collision cross sections relative to their non-vortex counterparts [22,48–60]. Recently, a few studies have been undertaken on molecular targets, which showed that like atomic targets, the TDCSs for ionization by vortex projectile were altered compared to TDCSs with plane wave projectiles [65–67]. For ionization of $H_2$, it was demonstrated that the incident vortex projectile's orbital angular momentum changed the interference pattern observed in the TDCSs [65].

Despite the many investigations of the role of electron vortex properties on collision cross sections, the idea of using vortex projectiles to study projectile coherence effects has only arisen in the last year [68]. The initial study of such effects was performed on an atomic target, where the observation of coherence effects is challenging. The use of a molecular target to study electron projectile coherence effects is ideal because the presence or absence of interference effects in the TDCSs is dependent upon on the projectile's coherence length. Thus, the ionization of molecular targets by electron vortex projectiles provides a straight-forward way to directly examine the role of electron projectile coherence. The results presented here predict that the coherent nature of electron vortex projectiles can be controlled through their intrinsic properties and that their coherence has a significant effect on the TDCSs.

The remainder of the paper is organized as follows. Section II contains details of the theoretical treatment. Section III presents the TDCSs for LG and Bessel projectiles. Section IV contains a summary of the work.

**II. Theory**

The TDCSs for electron vortex ionization of $H_2^+$ were calculated using the perturbative First Born Approximation (FBA) [48,49], which is applicable for the projectile energies and scattering geometries used here and is sufficient to capture the relevant features of the TDCS. In the FBA, the TDCS is proportional to the square of the transition matrix $T_{fi}^V$

$$\frac{d^3\sigma}{d\Omega_1 d\Omega_2 dE_2} = \mu_{pt}^2 \mu_{ie} \frac{k_f k_e}{k_i} |T_{fi}^V|^2, \tag{1}$$

where

$$T_{fi}^V = -(2\pi)^{3/2} <\Psi_f|V_i|\Psi_i^V>. \tag{2}$$

The reduced mass of the final state ion and ionized electron is $\mu_{ie}$ and the reduced mass of the projectile and target molecule is $\mu_{pt}$. The momenta are given by $\vec{k}_i$ for the incident projectile, $\vec{k}_f$ for the scattered projectile, and $\vec{k}_e$ for the ionized electron. Inserting complete sets of position states allows Eq. (2) to be written as an integral over all of position space for each of the particles in the collision. The projectile wave functions are expressed in cylindrical coordinates $(\rho_1, \varphi_1, z_1)$ and the bound and ionized electron wave functions are expressed in spherical coordinates $(r_2, \theta_2, \varphi_2)$. The origin is located at the target center of mass. Within this geometry, the projectile momenta can be written in terms of their respective longitudinal and transverse components as $\vec{k}_i = k_{i\perp}\hat{\rho}_{1i} + k_{iz}\hat{z}_1$ and $\vec{k}_f = k_{f\perp}\hat{\rho}_{1f} + k_{fz}\hat{z}_1$. We consider here the coplanar scattering geometry, in which the incident projectile, final projectile, and ionized electron momenta all lie in the same plane (x-z plane of Fig. 1a).

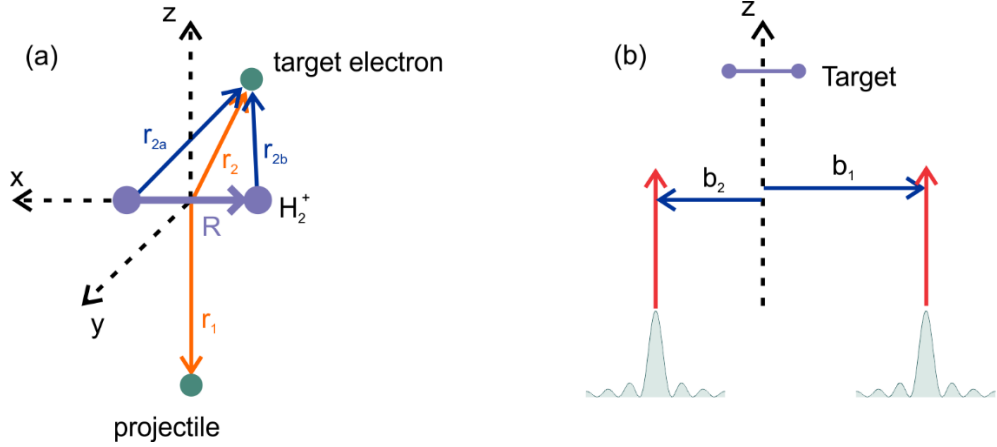

Figure 1 (a) Coordinate system for $e^- + H_2^+$ collision system. The incident projectile propagates along the z-direction and the target molecule is shown with an alignment perpendicular to the incident projectile beam. The orange vectors represent the position of the projectile ($\vec{r}_1$) and target electron ($\vec{r}_2$) relative to the center of mass of the target. The blue vectors represent the position of the target electron relative to each nuclear center ($\vec{r}_{2a}, \vec{r}_{2b}$). (b) Schematic of incident projectile wave packet impinging on a target atom. The shaded green region is the transverse profile of an incident Bessel projectile. Because the LG and Bessel twisted wave functions are not uniform in the transverse direction, an impact parameter $\vec{b}$ must be defined. Two possible values of $\vec{b}$ are shown ($\vec{b}_1, \vec{b}_2$), and the red arrows indicate the incident projectile's propagation direction.

The initial state wave function is expressed as a product of the incident vortex wave function $\chi_{\vec{k}_i}^V(\vec{r}_1)$ and the target molecular wave function $\Phi(\vec{r}_2, \vec{R})$

$$\Psi_i^V = \chi_{\vec{k}_i}^V(\vec{r}_1)\Phi(\vec{r}_2, \vec{R}), \tag{3}$$

where $\vec{R}$ is the internuclear vector (see Fig. 1a). The incident projectile vortex beam is chosen to be either a localized Laguerre-Gauss beam or a delocalized Bessel beam. Because the Bessel and LG vortex beams are non-uniform in the transverse direction, their transverse alignment relative to the target center of mass must be considered. The alignment of the projectile's transverse center with the target's center of mass is accounted for through the introduction of an impact parameter $\vec{b}$, such that $\vec{b}$ points transversely from the target center of mass to the center of the impinging vortex projectile wave packet (Fig. 1b).

For $\vec{b} = 0$, the wave function for the Bessel projectile is given by

$$\chi^V_{\vec{k}_i}(\vec{r}_1, \vec{b} = 0) = \chi^B_{\vec{k}_i,l}(\vec{r}_1, \vec{b} = 0) = \frac{e^{il\varphi_1}}{2\pi} J_l(k_{i\perp}\rho_1) e^{ik_{iz}z_1}, \tag{4}$$

where $J_l(k_{i\perp}\rho_1)$ is the Bessel function and $k_{i\perp}$ is the incident transverse momentum that can be written in terms of the opening angle $\alpha$

$$k_{\perp i} = k_i \sin\alpha. \tag{5}$$

Using a superposition of tilted plane waves [52], the Bessel wave function can be written as

$$\chi^B_{\vec{k}_i,l}(\vec{r}_1, \vec{b} = 0) = \frac{(-i)^l}{(2\pi)^2} \int_0^{2\pi} d\phi_{ki}\, e^{il\phi_{ki}} e^{i\vec{k}_i \cdot \vec{r}_1}. \tag{6}$$

Similarly, for $\vec{b} = 0$, the wave function for the LG projectile is given by [52]

$$\chi^V_{\vec{k}_i}(\vec{r}_1, \vec{b} = 0) = \chi^{LG}_{\vec{k}_i,l}(\vec{r}_1, \vec{b} = 0) = \frac{N}{w_0} e^{il\varphi_1} \left(\frac{\rho_1\sqrt{2}}{w_0}\right)^{|l|} L_n^{|l|}\left(\frac{2\rho_1^2}{w_0^2}\right) e^{-2\rho_1^2/w_0^2} \frac{e^{ik_{iz}z_1}}{\sqrt{2\pi}}, \tag{7}$$

where $N$ is a normalization constant[1], $w_0$ is the beam waist, and $L_n^{|l|}\left(\frac{2\rho_1^2}{w_0^2}\right)$ is an associated Laguerre polynomial with orbital angular momentum $l$ and index $n$ that is related to the number of nodes for a given $l$. For ease of computation, it is convenient to use the expression of the LG wave function as a convolution of Bessel functions over transverse momentum [52]

$$\chi^{LG}_{\vec{k}_i,l}(\vec{r}_1, \vec{b} = 0) = \frac{N}{\sqrt{2}} \frac{e^{il\varphi_1}}{n!} \int_0^\infty dk_{i\perp}\, e^{-w_0^2 k_{i\perp}^2/8} \left(\frac{w_0 k_{i\perp}}{\sqrt{8}}\right)^{2n+l+1} J_l(k_{i\perp}\rho_1) \frac{e^{ik_{iz}z_1}}{\sqrt{2\pi}}. \tag{8}$$

Combining equations (8) and (4), the LG wave function can be expressed as a convolution of Bessel projectile wave functions over transverse momentum

$$\chi^{LG}_{\vec{k}_i,l}(\vec{r}_1, \vec{b} = 0) = \frac{N\sqrt{\pi}}{n!} \int_0^\infty dk_{i\perp} e^{-\frac{k_{i\perp}^2 w_0^2}{8}} \left(\frac{k_{i\perp} w_0}{\sqrt{8}}\right)^{2n+l+1} \chi^B_{\vec{k}_i,l}(\vec{r}_1, \vec{b} = 0). \tag{9}$$

The initial state $H_2^+$ molecular wave function is written as a linear combination of atomic orbitals with each orbital having an effective nuclear charge of $Z_{eff}$ [69]

---

[1] The expression for $N$ in [52] is slightly incorrect. $N$ was calculated numerically to ensure normalization of the LG wave function.

$$\Phi(\vec{r}_2) = \frac{Z_{eff}^{\frac{3}{2}}}{\sqrt{2}} \left[ e^{-Z_{eff}|\vec{r}_{2a}|} + e^{-Z_{eff}|\vec{r}_{2b}|} \right]. \tag{10}$$

The coordinates $\vec{r}_{2a}$ and $\vec{r}_{2b}$ are the position vectors for the target electron relative to each of the nuclei (see Fig. 1(a))

$$\vec{r}_{2a} = \vec{r}_2 + \vec{R}/2$$

$$\vec{r}_{2b} = \vec{r}_2 - \vec{R}/2. \tag{11}$$

For an equilibrium separation of $|\vec{R}| = 2$ a.u., $Z_{eff} = 1.25$ yields the correct ionization potential of 29.9 eV.

The final state wave function is written as a product of the scattered projectile wave function $\chi_{\vec{k}_f}(\vec{r}_1)$, the ionized electron wave function $\chi_{\vec{k}_e}(\vec{r}_2)$, and the post-collision Coulomb interaction (PCI) $M_{ee}$

$$\Psi_f = \chi_{\vec{k}_f}(\vec{r}_1) \chi_{\vec{k}_e}(\vec{r}_2) M_{ee}. \tag{12}$$

Given the large projectile energy, it is sufficient to assume that the scattered projectile leaves the collision as a plane wave given by

$$\chi_{\vec{k}_f}(\vec{r}_1) = \frac{e^{i\vec{k}_f \cdot \vec{r}_1}}{(2\pi)^{3/2}}. \tag{13}$$

A two-center Coulomb wave [70] is used to model the ionized electron

$$\chi_{\vec{k}_e}(\vec{r}_2) = \frac{e^{i\vec{k}_e \cdot \vec{r}_2}}{(2\pi)^{\frac{3}{2}}} \Gamma(1 - i\eta) e^{-\frac{\pi\eta}{2}} {}_1F_1\left(i\eta, 1, -ik_e r_{2a} - i\vec{k}_e \cdot \vec{r}_{2a}\right) \Gamma(1 - i\eta) e^{-\frac{\pi\eta}{2}} {}_1F_1\left(i\eta, 1, -ik_e r_{2b} - i\vec{k}_e \cdot \vec{r}_{2b}\right), \tag{14}$$

where $\Gamma(1 - i\eta)$ is the gamma function and $\eta = Z_C Z_e/k_e$ is the Sommerfeld parameter with $Z_C$ the effective, screened nuclear charge seen by the ionized electron and $Z_e = 1$ is the charge of the electron. In the calculations presented here, we use $Z_C = 1$, which is the charge of each of the protons in the final state. The ejected electron wave function of Eq. (14) is orthogonalized to the target electron wave function of Eq. (10) through the Gramm-Schmidt procedure.

The Ward-Macek factor [71] is used to include the post-collision Coulomb repulsion between the two outgoing final state electrons

$$M_{ee} = N_{ee} \left| {}_1F_1\left(\frac{i}{2k_{fe}}, 1, -2ik_{fe}r_{ave}\right) \right|, \tag{15}$$

where

$$N_{ee} = \sqrt{\frac{\frac{\pi}{k_{fe}}}{k_{fe}\left(e^{\frac{\pi}{k_{fe}}} - 1\right)}}. \tag{16}$$

The relative momentum is $k_{fe} = \frac{1}{2}|\vec{k}_f - \vec{k}_e|$ and the average coordinate $r_{ave} = \frac{\pi^2}{16\epsilon}\left(1 + \frac{0.627}{\pi}\sqrt{\epsilon}\ln\epsilon\right)^2$, where $\epsilon = (k_f^2 + k_e^2)/2$ is the total energy of the two outgoing electrons.

The perturbation $V_i$ is the Coulomb interaction between the projectile and target molecule, which is given by

$$V_i = \frac{-1}{|\vec{r}_1 - \vec{R}/2|} + \frac{-1}{|\vec{r}_1 + \vec{R}/2|} + \frac{1}{r_{12}}. \tag{17}$$

Combining the above equations allows for the transition matrices for on-center Bessel and LG projectiles to be written in terms of the transition matrix for a non-vortex incident plane wave projectile $T_{fi}^{NV}$

$$T_l^B(\vec{b} = 0) = \frac{(-i)^l}{2\pi}\int_0^{2\pi} d\phi_{k_i} e^{il\phi_{k_i}} T_{fi}^{NV} \tag{18}$$

and

$$T_{l,n}^{LG}(\vec{b} = 0) = \frac{N\sqrt{\pi}}{n!}\int_0^\infty d\kappa \, e^{-\frac{w_0^2\kappa^2}{8}}\left(\frac{w\kappa}{\sqrt{8}}\right)^{2n+l+1} T_l^B \tag{19}$$

In a scattering experiment with a gas target, a specific impact parameter cannot be selected or controlled, and it is therefore necessary for theory to integrate over all possible impact parameters in order to provide an accurate comparison with possible experiments. For a Bessel projectile, the TDCS integrated over impact parameter is given by [48,50,56]

$$\frac{d^3\sigma_B}{d\Omega_1 d\Omega_2 dE_2}\bigg|_{int\ b} = \mu_{pt}^2 \mu_{ie} \frac{k_f k_e}{k_{iz}(2\pi)} \int_0^{2\pi} |T_{fi}^{NV}|^2 d\phi_{k_i}, \qquad (20)$$

and for a LG projectile, the TDCS integrated over impact parameter is given by [68]

$$\frac{d^3\sigma_{LG}}{d\Omega_1 d\Omega_2 dE_2}\bigg|_{int\ b} = \mu_{pt}^2 \mu_{ie} \frac{k_f k_e}{k_i} \frac{N^2 \pi}{(n!)^2 (2\pi)} \int_0^\infty \frac{e^{-\frac{k_{i\perp}^2 w_0^2}{4}}}{k_{i\perp}^2} \left(\frac{k_{i\perp}^2 w_0^2}{8}\right)^{2n+l+1} |T_{fi}^{NV}|^2 k_{i\perp} dk_{i\perp} d\phi_{k_i}. \qquad (21)$$

### III. Results

#### A. Controlling Transverse Coherence Length using Laguerre-Gauss Projectiles

The LG and Bessel projectiles provide different means through which to examine the role of projectile coherence and localization. For LG projectiles, the transverse coherence length $\delta$ can be defined using the quantum mechanical uncertainty $\Delta \rho$ of the transverse wave packet

$$\delta = 2\Delta\rho = [\langle \rho^2 \rangle - \langle \rho \rangle^2]^{1/2}, \qquad (22)$$

which varies linearly with the beam waist $\Delta \rho \propto w_0$. The factor of 2 is included to account for the presence of projectile density on both sides of the origin (i.e. when the azimuthal angle is 0 or $\pi$). The projectile's transverse coherence length can be directly controlled through the beam waist. In addition, the transverse coherence length changes with the Laguerre-Gauss indices $n$ and $l$ (from now on labeled as $(n,l)$). A few sample values of the uncertainty for LG projectiles are listed in Table 1 and the projectile transverse densities are plotted in Fig. 2. For comparison, the internuclear separation for $H_2^+$ is 2 a.u., and it is expected that interference features will be present in the TDCSs when $\delta \gtrsim 2$ a.u. For (1,0) LG projectiles, the transverse density contains more than one peak, and Table 1 includes the coherence length calculated for all peaks, as well as for only the primary (largest) peak.

|  | δ all peaks (0,0) | δ all peaks (1,0) | δ primary peak (1,0) | δ all peaks (0,1) |
|---|---|---|---|---|
| $w_0 = 0.5$ a.u. | 0.24 | 0.32 | 0.15 | 0.24 |
| $w_0 = 1$ a.u. | 0.46 | 0.46 | 0.30 | 0.48 |
| $w_0 = 4$ a.u. | 1.86 | 2.52 | 1.24 | 1.94 |
| $w_0 = 8$ a.u. | 3.70 | 5.04 | 2.48 | 3.86 |

*Table 1* Transverse coherence length $\delta$ of LG projectiles in atomic units calculated using Eq. (22). The LG parameters $n$ and $l$ are labeled as $(n,l)$. For $(1,0)$ LG projectiles, $\delta$ is listed for the case when all peaks are included in the density, as well as for when only the primary peak is included.

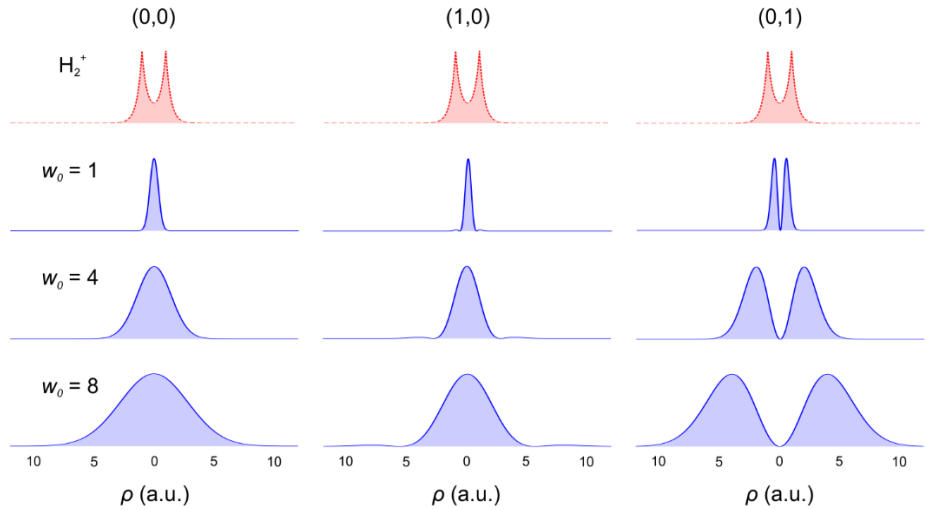

Figure 2 Transverse profiles of the $H_2^+$ electron density (top row, red dashed line) and the LG beam (rows 2-4, blue solid line) as a function of transverse distance $\rho$. The LG parameters $n$ and $l$ are labeled at the top for each column $(n,l)$ and the beam waist is labeled at the left for each row ($w_0$). All profiles are normalized to 1 to provide a qualitative comparison of width. The target $H_2^+$ electron density is the same for each column.

Figure 3 shows the coplanar TDCSs for ionization of $H_2^+$ by LG projectiles for different $l$ and $n$ as a function of ejected electron angle and beam waist. The TDCSs have been averaged over impact parameter. The color indicates magnitude of the TDCS (cooler colors are smaller TDCSs and warmer colors are larger TDCSs). Two fixed orientations of the $H_2^+$ molecule are shown relative to the incident beam direction. The first column shows TDCSs for $H_2^+$ aligned parallel to the beam direction and the second column shows TDCSs when the molecule is aligned perpendicular to the beam direction. The third column shows the TDCSs averaged over all possible orientations. The white horizontal dashed line represents the beam waist value that yields a transverse coherence length of $\delta = 2$ a.u. (i.e. equal to the internuclear separation).

The bottom row of each subpanel in Fig. 3 is a color bar that contains the TDCS for ionization of $H_2^+$ by a plane wave projectile (see label in Fig. 3(c)). In this case, the projectile is completely delocalized and the TDCSs show clear interference features due to ionization from the two nuclear centers. The top row of each subpanel in Fig. 3 shows a color bar that contains the TDCS for ionization of atomic H by a LG projectile with $w_0 = 0.4$ a.u. (see label in Fig. 3(c)). In this case, the projectile is localized, and no interference effects are present due to the single ionization center of the atomic target.

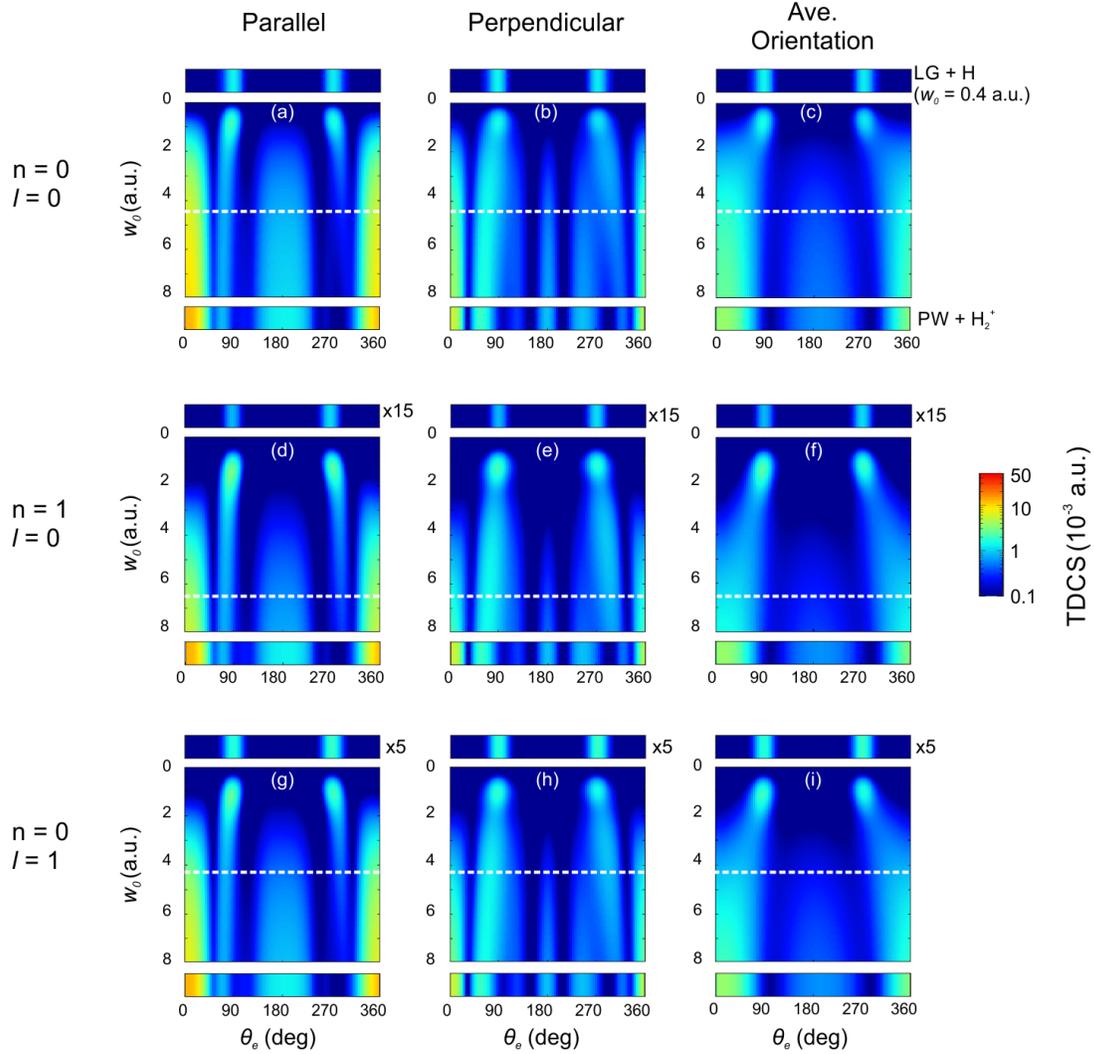

Figure 3 TDCSs for ionization of $H_2^+$ by LG projectiles as a function of beam waist ($w_0$) on the vertical axis and ejected electron angle ($\theta_e$) on the horizontal axis. The magnitude of the TDCS is indicated by color. Sub-panels (a) – (c) show TDCSs for $n = 0, l = 0$ (0,0); (d) – (f) for $n = 1, l = 0$ (1,0); (g) – (i) for $n = 0, l = 1$ (0,1). In the first column, the molecule's axis is aligned parallel with the incident beam direction. In the second column, the molecule's axis is aligned perpendicular to the incident beam direction. In the third column, the TDCSs were averaged over all molecular axis orientations. The horizontal dashed white line indicates the beam waist value that yields a transverse coherence length of $\delta = 2$ a.u. (i.e. equal to the internuclear separation). Within each sub-panel, the top row is the TDCS for ionization of atomic hydrogen by a LG projectile with $w_0 = 0.4$ a.u. (as indicated in (c)). The atomic hydrogen TDCSs have been multiplied by 15 (d)-(f) and 5 (g)-(i). The bottom row is the TDCS for ionization of $H_2^+$ by a plane wave (as indicated in (c)).

For ionization of $H_2^+$ by a LG projectile with a small beam waist, the TDCSs show two, narrow peaks at $\theta_e = 90°$ and $270°$, independent of molecular orientation. The shape of these TDCSs is similar to the TDCS for ionization of H by a narrow LG projectile (color bar above the subpanels), which confirms that projectiles with small transverse coherence length cause ionization from only one of the nuclear centers.

As the beam waist increases, these two peaks shift slightly toward the forward direction, and an additional peak at $\theta_e = 0°$ appears. As the beam waist is further increased, a peak at $\theta_e = 180°$ appears and the TDCSs then show the same interference structures as the TDCS for a plane wave projectile. The presence of the full pattern of interference structures in the LG TDCSs occurs approximately at a beam waist value that corresponds to a projectile transverse coherence length equal to the internuclear separation. This indicates that when the transverse coherence length is greater than or equal to the internuclear separation, ionization can occur from either nuclear center with the amplitudes interfering. The value of the beam waist that results in a transverse coherence length equal to the internuclear separation can be interpreted as a threshold value for the coherence of the electron projectile. Beyond the threshold value, the interference structures persist in the TDCSs with very little change which is consistent with the requirement that the transverse coherence length must be large enough to perturb the target electron wave function at both nuclear centers in order for the ionization amplitudes to interfere.

The onset of interference features in the TDCSs for (1,0) LG projectiles approximately coincides with the beam waist threshold found for the coherence length of the primary peak and not all peaks, indicating that the small side peak for the (1,0) LG projectile does not significantly influence the TDCSs and that ionization primarily occurs due to the main peak. As shown in Table 1 and Fig. 2, for a given beam waist, the transverse coherence length of the LG beam for (0,0) is larger than that of the primary lobe of a LG beam with (1,0), and this results in the interference effects being observed at a smaller beam waist value for (0,0) LG projectiles (Fig. 3(a)-(c)) than for (1,0) LG projectiles (Fig. 3(d)-(f)).

For (0,1) LG projectiles (Fig. 3(g)-(i)), the LG transverse profile contains two equal lobes with a node at the origin. The width of this node increases with increasing beam waist, and its presence is caused by the non-zero orbital angular momentum of the (0,1) LG projectile. The TDCSs for (0,1) LG projectiles are very similar to those of (0,0) LG projectiles, and the threshold beam waist value that leads to a coherent projectile is also similar. This indicates that for the kinematics considered here, the orbital angular momentum of the projectile plays a small role. Additionally, the double peaked structure of the projectile's transverse density for (0,1) LG projectiles does not alter the TDCSs compared to those of the (0,0) LG projectile. This is likely due to the uncertainty in the alignment of the projectile and target for the TDCSs presented in Fig. 3 (i.e. the average of the TDCSs over impact parameter). The inclusion of TDCSs for projectiles with non-zero impact parameters introduces contributions to the TDCSs from additional orbital angular momenta [56] and has the effect of washing out orbital angular momentum-specific features in the TDCSs.

Some minor qualitative differences are observed in the TDCSs for large beam waist ($w_0 = $ 8 a.u.) relative to the plane wave TDCSs. These are most noticeable when the molecular axis is perpendicular to the beam direction, in which case more individual peaks are observed in the plane wave TDCSs than the LG TDCSs. These differences are due to the localization of the LG projectile, despite the large beam waist. When the beam waist was increased to $w_0 \geq 20$ a.u. (not shown), the TDCSs were identical to the plane wave TDCSs.

The TDCSs averaged over molecular orientation show fewer interference features because any orientation-dependent effects are averaged out. However, the delineation in the shape of the TDCSs for coherent and incoherent projectiles is still observable. At small beam waist, the TDCSs exhibit only the two peaks at 90° and 270° that are characteristic of ionization from a single nuclear center. As the beam waist increases, these peaks broaden and become a single forward peak centered at 0°. This shift and broadening occurs at a smaller beam waist value for the (0,0) LG projectile and at larger beam waist value for the (0,1) and (1,0) LG projectiles, consistent with the dependence of the transverse coherence length on beam waist for the different LG parameters. For beam waists larger than the threshold for a coherent projectile, a peak is observed at 180° in the TDCSs as a result of interference from two nuclear centers.

**B. Controlling Localization with Bessel Projectiles**

For the Bessel projectile, the transverse coherence as defined in Eq. (22) is infinite, but an effective width (or transverse localization) of the wave packet can be defined. As the projectile opening angle increases, the transverse momentum increases and the individual peaks in the projectile transverse density become narrower (see Fig. 4). In the limit that $\alpha = 0$ and $l = 0$, the Bessel projectile is identical to a completely delocalized plane wave. Thus, the opening angle provides a mechanism by which to change the effective transverse localization of the Bessel projectile. To quantify the localization, we assume that the dominant contribution to ionization comes from the primary (largest) peak of the Bessel wave function, and we define the transverse localization of the Bessel projectile as the transverse uncertainty calculated using only the density up to the first zero. A similar assumption led to accurate predictions of coherence length for the (1,0) LG projectile. A few values of the transverse localization for the Bessel projectile are listed in Table 2.

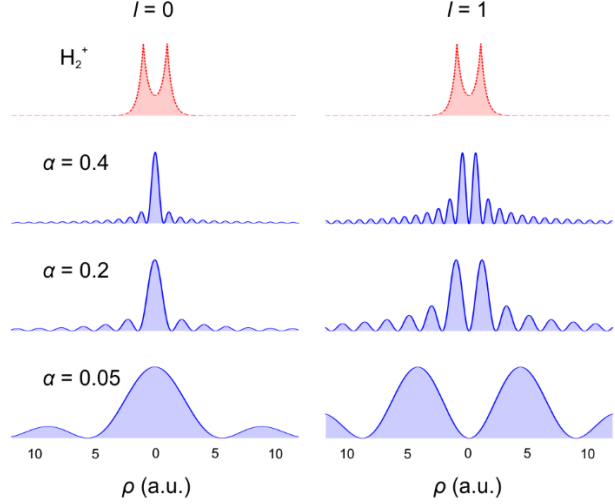

Figure 4 Transverse profiles of the $H_2^+$ electron density (top row, red dashed line) and the Bessel beam (rows 2-4, blue solid line) as a function of transverse distance $\rho$. The Bessel projectile orbital angular momentum is labeled at the top for each column ($l$) and the opening angle is labeled at the left for each row ($\alpha$). All profiles are normalized to 1 to provide a qualitative comparison of width. The target $H_2^+$ electron density is the same for each column.

|  | $\delta$ primary peak $l = 0$ | $\delta$ primary peak $l = 1$ |
|---|---|---|
| $\alpha = 0.4$ a.u. | 0.28 | 0.38 |
| $\alpha = 0.2$ a.u. | 0.56 | 0.76 |
| $\alpha = 0.1$ a.u. | 1.10 | 1.50 |
| $\alpha = 0.05$ a.u. | 2.20 | 3.00 |

*Table 2* Transverse localization $\delta$ of Bessel projectiles in atomic units for projectiles with orbital angular momenta of $l = 0,1$.

Figure 5 shows the coplanar TDCSs for ionization of $H_2^+$ by a Bessel projectile as a function of ejected electron angle and opening angle. As in Fig. 3, the color indicates magnitude of the TDCS, and two fixed orientations of the $H_2^+$ molecule are shown in addition to the TDCSs averaged over all possible orientations. The TDCSs have been averaged over impact parameter, which results in contributions from all orbital angular momenta of the projectile contributing to the TDCS [56]. Thus, for Bessel projectiles averaged over impact parameter, a specific orbital angular momentum of the projectile cannot be identified, although prior work on the ionization of atomic targets has shown that the dominant contributions to the TDCSs are from low orbital angular momentum terms [48,50,65].

The top row of each subpanel in Fig. 5 is a color bar that contains the TDCS for ionization of $H_2^+$ by a Bessel projectile with $\alpha = 0.4$ rad (see label in Fig. 5(c)). In this case, the projectile is localized, and no interference effects are present due to the single ionization center of the atomic target. A trace along the horizontal axis of each subpanel ($\alpha = 0$) yields the plane wave TDCS in which the projectile is completely delocalized. Here, clear interference features are visible due to ionization from the two nuclear centers.

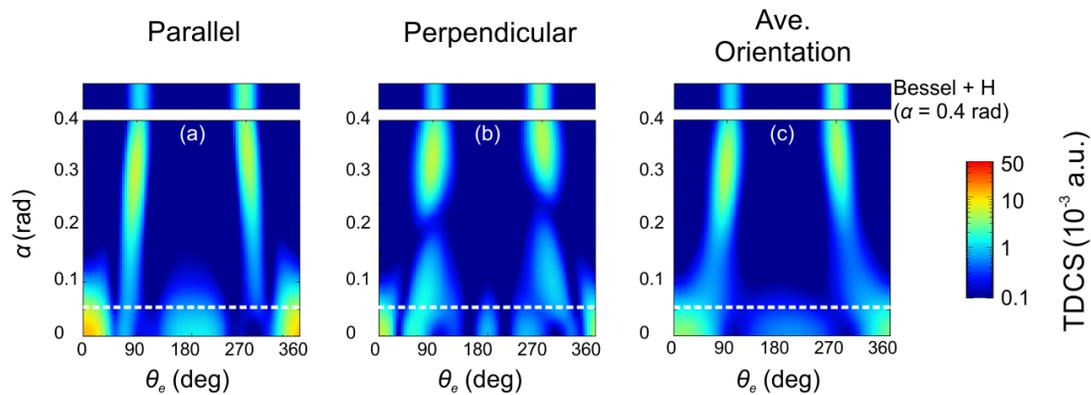

Figure 5 TDCSs for ionization of $H_2^+$ by Bessel projectiles as a function of opening angle ($\alpha$) on the vertical axis and ejected electron angle ($\theta_e$) on the horizontal axis. The magnitude of the TDCS is indicated by color. Sub-panel (a) is the TDCS for the molecule's axis is aligned parallel with the incident beam direction. Sub-panel (b) is the TDCS for the molecule's axis is aligned perpendicular to the incident beam direction. In sub-panel (c), the TDCSs were averaged over all molecular axis orientations. The horizontal dashed white line indicates the opening angle value that yields a transverse localization of $\delta = 2$ a.u. (i.e. equal to the internuclear separation). Within each sub-panel, the top row is the TDCS for ionization of atomic hydrogen by a Bessel projectile with $\alpha = 0.4$ rad (as indicated in (c)). A trace along the x-axis for $\alpha = 0$ yields the TDCS for ionization of $H_2^+$ by plane wave.

As shown in Fig. 4 and Table 2, Bessel projectiles with large opening angles have a highly localized main peak in their transverse profile and a small transverse localization. At large opening angles, the TDCSs for ionization of $H_2^+$ resemble those of ionization of atomic hydrogen by a localized projectile, independent of molecular orientation. This demonstrates that Bessel projectiles with large opening angles behave as localized wave packets and are only able to cause ionization from one of the nuclear centers.

As the opening angle decreases, the Bessel projectile's transverse density broadens and becomes less localized. Much like with LG projectiles, the peaks in the TDCSs shift toward the forward direction as the projectile transverse density broadens, and below a threshold value of opening angle, interference structures are observed in the TDCSs. The horizontal white dashed lines in Fig. 5 show the opening angle that yields a transverse localization of the $l = 0$ Bessel projectile of 2 a.u. Fig. 5 shows that for opening angles less than this threshold value, the projectile's transverse profile is broad enough to cause ionization from both nuclear centers, resulting in an interference pattern consistent with ionization from two nuclear centers. Because the threshold value of opening angle yields a Bessel projectile transverse localization that coincides with the internuclear separation of the target, we can conclude that despite the Bessel projectile's infinite transverse coherence length, it behaves as a localized wave packet. Additionally, the threshold value of the opening angle calculated using only the main peak of the $l = 0$ projectile density qualitatively agrees with the appearance of the interference features in the TDCSs. This is an indication that ionization predominantly occurs due to the main peak in the projectile density and supports our use of only this peak to estimate projectile coherence. It also indicates that the lowest orbital angular momentum value is the dominant contribution to the TDCSs, even though contributions from all angular momenta are included.

Overall, the TDCSs for LG and Bessel projectiles exhibit qualitatively similar behavior as either beam waist or opening angle is changed. At small beam waist and large opening angle, the TDCSs resemble those for ionization of atomic hydrogen, indicating a highly localized and incoherent projectile. In contrast, at large beam waist and small opening angle, the TDCSs resemble those for ionization of $H_2^+$ by a plane wave projectile, indicating a coherent projectile. Thus, projectile coherence can be controlled for both LG and Bessel projectiles through the choice of the wave packet parameters.

## IV. Summary

In the last decade, much attention has been focused on the role of projectile coherence and localization in heavy particle atomic and molecular collisions, and it has been shown that theoretical models must consider the transverse coherence length of the projectile when calculating collision cross sections. However, until recently, no studies have examined the role of projectile coherence in electron-impact collisions because it has been thought that the deBroglie wavelength of an electron projectile is sufficiently large to ensure that the projectile is fully coherent. We have demonstrated that it is possible to alter an electron projectile's coherence by using vortex projectiles and that the projectile's transverse coherence length has a significant and observable effect on the collision cross sections.

We presented theoretical TDCSs for electron-impact ionization of $H_2^+$ using Laguerre-Gauss and Bessel projectiles. We controlled the projectile's transverse coherence length and localization through the parameters of the vortex projectile, such as beam waist and opening angle. The use of a projectile with a small beam waist or large opening angle yielded TDCSs that resembled those found for ionization of atomic hydrogen, indicating that the projectile was localized and that ionization occurred from only one of the nuclear centers of the molecule. In contrast, a projectile with a large beam waist or small opening angle yielded TDCSs that resembled those found for non-vortex, plane wave projectiles, and showed an interference pattern indicative of ionization from both nuclear centers.

We identified threshold values of beam waist and opening angle that separated coherent, delocalized projectiles from incoherent, localized projectiles. These threshold values coincided with the calculated coherence length and localization that equaled the molecule's internuclear separation, consistent with the premise that interference features in the TDCSs are observable when the projectile's transverse coherence length is greater than or equal to the internuclear separation. The presence and or absence of interference features in the TDCSs provided a clear indicator of projectile coherence, and our results demonstrated that the coherence of an electron projectile can be controlled and studied using twisted wave packets.

**Acknowledgments**

We gratefully acknowledge the support of the National Science Foundation under Grant No. PHY-1912093.